# Quantum-coupling between closely-spaced surfaces via transverse photons

K. P. Sinha[a], A. Meulenberg[b], P. L. Hagelstein[c]

[a] Research Laboratory of Electronics, Massachusetts Institute of Technology, Cambridge, MA 02139    Presently, Department of Physics, IISc, Bangalore 560012, India (email: kpsinha@gmail.com)

[b] The Charles Stark Draper Laboratory, Inc., Cambridge, MA 02139    Presently, Visiting Scientist - Department of Instrumentation, IISc, Bangalore 560012, India  (email: mules33@excite.com)

[c] Research Laboratory of Electronics, Massachusetts Institute of Technology, Cambridge, MA 02139

A quantum-mechanical formulation of energy transfer between closely-spaced surfaces is given. Coupling between the two surfaces arises from the atomic dipole-dipole interaction involving transverse-photon exchange. The exchange of photons at resonance enhances the radiation transfer. The spacing (distance) dependence is derived for the quantum well - quantum well situation. The interaction between two quantum wells, separated by gap $\tilde{l}$ (length, $\tilde{l}$, is dimensionless), is found to be proportional to $(1/\tilde{l}^4)\, T_{12}(\tilde{l})$, where $T_{12}(\tilde{l})$ is the radiation tunneling factor for the evanescent waves. Expressions for the net power transfer, in the near-field regime, from hot to cold surface for this case is given and evaluated for representative materials.

INTRODUCTION

Thermophotovoltaics (TPV), which converts light (more accurately thermally-generated, long-wavelength-infrared radiation) from a heated surface into electricity, is beginning to grow as a result of new materials capabilities. Specifically, new semiconductor materials allow the photoconverters to convert long-wavelength light into electrical power more efficiently by providing a narrower electrical band gap, better generation of photo-excited minority carriers and their collection at the p-n junction, and reduced recombination dark-current, which controls the open-circuit voltage of the devices. However, there are some approaches, based on improved technical capabilities, which can compound all of these improvements and, potentially, allow TPV to become even more useful.

It has been demonstrated recently that, even for practical-size devices (2x2 cm), the optical throughput is greatly enhanced in the case of a very narrow vacuum gap.[1] This enhancement, through the use of a "submicron" gap, is the basis for Microgap Thermophotovoltaics (MTPV).[2,3]

A requirement of a TPV system is a thermal difference between the emitter and PV device. This means that thermal isolation in the form of distance (or of a vacuum, if the distance is small) must separate the devices. However, it is known that elimination of the gap can greatly enhance the transfer of radiation between them. The ability to increase the transfer of optical photons across a very small gap (a fraction of a wavelength) without allowing heat flow, via phonons, has also been demonstrated.[4]

One enhancement mechanism, available to MTPV and demonstrated in Ref. 1, is purely a physical-optics effect. It involves the development of an "effective" refractive index ($n_g$) within the microgap. As the gap approaches zero width, its effective refractive index approaches that of the emitter and PV device (assume $n_e = n_{pv}$ for optimum results in MTPV). As the difference between refractive indices of the devices and the gap diminishes, the critical angle of total internal reflection (TIR) increases and the percentage of thermally-generated light that can escape from the emitter increases. As the gap width approaches zero, the enhancement from this mechanism approaches a maximum of $n_e^2$. This effect (called the $n^2$ effect) is a macroscopic effect and is independent of the distance between <u>individual</u> radiators (excited atoms or dipoles) and absorbers (ground-state atoms) in the emitter and PV device, respectively.

However, the subject of this paper is enhancement of the optical coupling between atoms that <u>is</u> dependent on the distance between <u>individual</u> radiators and absorbers. This process will be called "resonance-enhancement." ("Proximity-enhancement" includes both the $n_e^2$ and the resonant-enhancement effects.)

Resonance enhancement involves the interaction of dipole oscillators that are close enough together to be "coupled" by a photon. This "correlated" interaction provides enhanced optical coupling and may be compared with the non-correlated interaction of independent creation and absorption of photons. Closer dipoles (in space and in transition energy or frequency) give stronger coupling. Thus, the phenomenon is both a proximity effect and a resonance effect.

In the following, the derived expression will explicitly contain the three effects, namely, the gapwidth and source/absorber-separation dependence, the resonance enhancement, and the $n^2$ enhancement. The quantum-coupling approach presented here, in the context of small-gap TVP converters, has not been considered previously for macroscopic devices. There do exist, however, relevant papers on the radiation field and fluctuations in close proximity to a hot emitter, for example, those based on a classical radiation-intensity approach[4] and others based on a classical formulation of fluctuation electrodynamics.[5] Most of the published studies[6,7,8,9,10], follow the latter approach in which the randomly fluctuating charge and current densities provide a source of radiation.

THEORETICAL MODEL

In the current model, we consider two-level systems between separated dielectrics at different temperatures, as shown in Figure 1. The two slabs are separated by a dielectric gap[11] of width $l$, which can be varied. While our discussion has focused above on the case of coupling across a vacuum gap, it is possible that the first experiments to confirm the model will be done with a thin solid gap made of an insulator, and the thermal excitation of the emitter will be replaced by an optical excitation scheme. Various parameters of the regions, such as temperatures, $T_1$ ($>T_2$), dielectric functions, $\varepsilon_i(\omega)$ - where $\omega$ is the frequency - will be denoted by the corresponding suffixes.

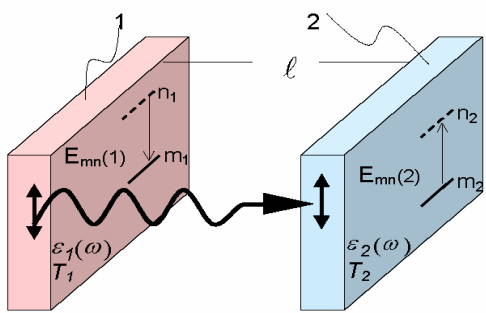

*Figure 1. Coupling between two-level quantum systems separated by a gap.*

We recognize that there may be many two-level systems on both the emitter side and on the absorber side of the gap. We will assume that the coupling between the two-level systems is radiative, including contributions from both electrostatic (Coulomb) and electrodynamic fields. This implies an underlying Hamiltonian of the form

$$\hat{H} = \sum_i \hat{H}_{atom}^{(1)}[i] + \sum_j \hat{H}_{atom}^{(2)}[j] + \hat{H}_{rad} + \hat{H}_{coul} + V_{dipole} \qquad (1)$$

In this equation, the first summation is over emitter-side two-level atomic models indexed by $i$. The second summation is over the absorber-side two-level atomic models, indexed by $j$. The transverse radiation field is denoted through $\hat{H}_{rad}$, and the Coulomb interaction is included through $\hat{H}_{coul}$. The model involving the longitudinal electrodynamic mode of $H_{Coul}$ has been discussed by the authors in another paper.[12] In what follows, we consider coupling through $V_{dipole}$, which involves the transverse electric component of the Coulomb source field.

The atomic transitions are presumed to be dominated by the electric-dipole transitions. Thus, we may extract the dipole-dipole interaction from a multipole expansion of the interaction:

$$V_{dipole} = -\sum \boldsymbol{\mu}_j \cdot \mathbf{E}_T(\mathbf{r}_j) \qquad (2)$$

where $\boldsymbol{\mu}$ is the dipole-moment operator and $\mathbf{E}_T$ is the dipole (transverse-) electric-field operator.

Explicitly in the case of two dipoles in free space, the interaction Hamiltonian, $H_{int} = \mu(1) \cdot \mathbf{E}_{T2}(R_1) - \mu(2) \cdot \mathbf{E}_{T1}(R_2)$, where $\mathbf{E}_{T2}(R_1)$ and $\mathbf{E}_{T1}(R_2)$ are the Coulomb source fields (field at $R_1$ from dipole 2 and field at $R_2$ from dipole 1, respectively). The coupling between a pair of atoms (dipoles), on using the above interaction, will appear in fourth-order perturbation theory. However, it is convenient to transform the Hamiltonian by using a unitary transformation,[13,14,15,16] $H_t = e^{iS} H e^{-iS}$, where S is an operator chosen to eliminate the linear term $\mu \cdot \mathbf{E}_T$. Then, the effective interaction between two dipoles is obtained in the second-order perturbation theory.

Averaging over all polarizations and angles leads to the effective interaction between two <u>randomly-oriented</u> dipoles at a distance $R_{ij}$ apart in free space. For the near-field (NF) limiting form, this is:

$$[U^{NF}(R_{ij})]^{(2)} \approx -\frac{2}{3} \frac{|\mu_{mn}(2)|^2 |\mu_{mn}(1)|^2}{R_{ij}^6 \Delta E_{12}} \qquad (3)$$

where $\Delta E_{12}$, is a positive number and $\mu_{mn}$ are the matrix elements of the transition-dipole moments. In the near-field region, the dipoles are separated by much less than a wavelength at the energy of interest, $R_{ij} \ll \lambda$. In the far-field region, $R_{ij} \gg \lambda$.

For real-photon (or resonant) exchange, one side must be in the excited state and the other must be in the ground state. Thus, $\Delta E_{12}(r) = \hbar(\omega_2 - \omega_1)$, where $\omega_2$ and $\omega_1$ are frequencies associated with the respective transitions (Figure 2).

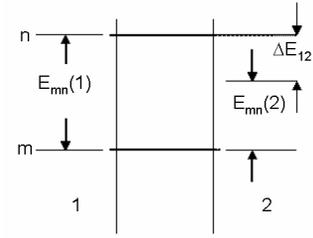

*Figure 2. Schematic representation of the two-medium, two-level system used in the model.*

For off-resonant photon exchange, $\Delta E_{12}$(o-r) is not equal to zero and the off-resonant result is found from the Cauchy Principal Value of the integral. For resonant-photon exchange, $\Delta E_{12} = \||E_{mn}(1)| - |E_{mn}(2)|\|$ and is vanishingly small, hence, the above equation must be solved with special care via the Method of Residues or use of the delta function, $\delta(\omega-\omega_0)$. Generally, the off-resonant terms are found to be small relative to the resonant terms.

For the London interaction,[14] the term in the denominator of Eq. (3) is $E_{mn}(1) + E_{mn}(2)$ rather than $\Delta E_{12}$. Since $E_{mn}(1) + E_{mn}(2) \gg \Delta E_{12} = \||E_{mn}(1)| - |E_{mn}(2)|\|$, any energy exchange via this pathway is negligible relative to real photon exchange.

In the system considered here, the absorbers are anchored in a medium, and hence, the complex dielectric constants of the material medium ($\varepsilon_i(\omega) = \varepsilon_i'(\omega) + i \varepsilon_i''(\omega)$) are involved and often appear in the corresponding equations (e.g., Van Kampen et al.[17] and Dzyaloshinskii et al.[18]). Although the dipoles are embedded in a dielectric, the transition matrix elements of the dipole moment are estimated from the oscillator strengths[12,18,19] (which are proportional to the absorption coefficients, as described by the imaginary portion of the dielectric function, or proportional to the square of the dipole matrix elements). In our case, we are using the dipole moments derived from the oscillator strengths, and therefore, Eq. (3) (with dielectric medium effects using the method of images) becomes:

$$[U^{NF}(R_{ij})]^{(2)} \approx -\frac{2}{3} \frac{16 \, \varepsilon_o(\omega)^2 |\mu_{mn}(2)|^2 |\mu_{mn}(1)|^2}{R_{ij}^6 |\varepsilon_1(\omega) + \varepsilon_o(\omega)|^2 |\varepsilon_2(\omega) + \varepsilon_o(\omega)|^2 \Delta E_{12}} \qquad (4)$$

The above formulation is for a pair of atoms (each a dipole), one on side 1 and another on side 2. The effect of <u>all</u> atoms on both sides can be obtained by integrations over both half spaces (see Ref. 15). If the single atom on side 1 is replaced by half space $Z \leq 0$ with $N_1$ atoms per unit volume (p.u.v.) and the other at half space $Z \geq l$ (where $l$ is the distance between the two slabs) with $N_2$ atoms (p.u.v.), the NF interaction energy per unit area for the current situation is:

$$[U^{NF}(l)]^{(2)} = -\frac{4}{9} \frac{(2\pi) \, N_1 N_2 \, \varepsilon_o(\omega)^2}{|\varepsilon_1(\omega) + \varepsilon_o(\omega)|^2 |\varepsilon_2(\omega) + \varepsilon_o(\omega)|^2} \frac{|\mu_{mn}(2)|^2 |\mu_{mn}(1)|^2}{\Delta E_{12}} \, G(l), \qquad (5)$$

where $G(l) = 1/l^2$. We have derived the geometric term, $G(l)$ for two quantum wells of width $W_{qw}$, separated by a vacuum gap $l$.

$$G = \frac{1}{l^2}\left[1 - \frac{2}{(1+(W_{qw}/l))^2} + \frac{1}{(1+(2W_{qw}/l))^2}\right] \quad (6)$$

A real, transverse photon is created on the hot side and absorbed in the cold side. As we are considering very short distances here (NF = near field), the evanescent[20] mode will dominate the radiation transfer in this region. Within the dipole-dipole interaction via the radiation field, the transfer of energy from slab 1 to slab 2 and the reverse process is to be considered. Before writing the full expression, we must note the following situations.

The probability of emission of a quantum of radiation depends on the factor $[n(\omega, T_1) +1]$ and the absorption on the other side depends on $n(\omega, T_2)$, where $n(\omega, T) = [1/\{\exp(\hbar \omega/K_B T) - 1\}]$, the Bose distribution function, $F(B)$. The expression will involve:

$$[n(\omega, T_1) +1][n(\omega, T_2)] - [n(\omega, T_2) +1][n(\omega, T_1)] = -[n(\omega, T_1) - n(\omega, T_2)] \quad (7)$$

For the case 1 quantum well on each side of the gap (areal structures), the net power transfer is the photon energy times the transition probability of photon transfer, $\Gamma = \Gamma_{12} - \Gamma_{21}$

$$P_{NF}(aa) = \int \hbar\omega \Gamma d\omega = \frac{2\pi}{\hbar} \int |U_{NF}(\omega,\bar{l})|^{(2)} \rho^{\alpha}(\omega) [n(\omega,T_1) - n(\omega,T_2)](\hbar\omega) d\omega \quad (8)$$

where $\rho^{\alpha}(\omega)$ is the areal density of states and $(\bar{l})$ is the dimensionless version of the gap width. (Explicitly, $(\bar{l}) = l\, n_g\, \omega/2\pi c$). Equation (8) may be expanded into:

$$P_{NF}(aa) = \frac{2\pi}{\hbar} \int (8\pi/9) \frac{|\mu_{mn}^{(1)}|^2 |\mu_{mn}^{(2)}|^2 N_1^a N_2^a}{|\hbar(\omega-\omega_o)|} (\hbar\omega) 4\pi \left(\frac{\omega\, n_e^2}{c^2}\right)\left(\frac{\omega^2 n_g^2}{(2\pi c)^2}\right)$$
$$* [n(\omega,T_1) - n(\omega,T_2)] F(M) G(\bar{l}) T_{12}(\bar{l})\, d\omega \quad (8a)$$

$$= \left(\frac{2\pi}{\hbar}\right)\int\left(\frac{8\pi}{9}\right)|\mu_{mn}^{(1)}|^2 |\mu_{mn}^{(2)}|^2 N_1^a N_2^a\, \delta(\omega - \omega_o)\left(\frac{\omega^4 n_i^4}{\pi c^4}\right)$$
$$* [n(\omega,T_1) - n(\omega,T_2)] F(M) G(\bar{l}) T_{12}(\bar{l})\, d\omega \quad (8b)$$

where $N_1^a$, $N_2^a$ are the number of dipoles per unit area on either side of the gap. Since, for very small gap size, the system resembles an effective uniform dielectric rather than an interrupted dielectric, we use $n_i$ for the (constant) refractive indices of the two materials and gap in question.

Integration over $\omega$ with use of the delta function leads to:

$$P_{NF}(aa) = \left(\frac{2\pi}{\hbar}\right)\left(\frac{8\pi}{9}\right)|\mu_{mn}(1)|^2 |\mu_{mn}(2)|^2 N_1^a N_2^a \left(\frac{\omega_o^4 n_i^4}{\pi c^4}\right)$$
$$* [n(\omega_o,T_1) - n(\omega_o,T_2)] T_{12}(\bar{l})\, G(\bar{l})\, F(M))$$
$$= (2\pi/\hbar)\ F(D)\ F(B)\ F(M)\ G(\bar{l})\ F(\omega_o)\ T_{12}(\bar{l}) \quad (9)$$

The factor, $F(\omega_o) = \left(\frac{\omega_o^4 n_i^4}{\pi c^4}\right)$, comes from the combined effect of frequency-dependent factors (namely areal density of states) and the energy (frequency) of the quanta being exchanged. The Bose factor, $F(B) = [n(\omega_o,T_1) - n(\omega_o,T_2)]$, provides the proportion of radiated quanta from each side.

We now define the other factors occurring in Eq. (9). The factor containing dipole terms is:

$$F(D) = \left(\frac{8\pi}{9}\right)|\mu_{mn}^{(1)}|^2 |\mu_{mn}^{(2)}|^2 N_1^a N_2^a \quad (10)$$

The factor containing the material dielectrics is:

$$F(M) = \frac{\varepsilon_0^2}{|\varepsilon_0 + \varepsilon_1|^2 |\varepsilon_0 + \varepsilon_2|^2}, \quad \text{the material factor} \tag{11}$$

The real part of the dielectric constant has only weak frequency dependence and is considered to be constant for our application.

The factor that gives the tunneling of evanescent waves across the gap (refractive index $n_g$) is:

$$T_{12}(\bar{l}) = \frac{1}{\left|1 + \frac{\left(|n_1|^2 + |n_g|^2\right)^2}{4|n_1|^2 |n_g|^2} \sinh^2(\bar{l})\right|} \tag{12}$$

$$G(\bar{l}) = 6\left(1/\bar{l}^4\right)\left(\frac{W_{qw}}{\lambda_g}\right)^2 \quad \text{and} \quad \bar{l} = \frac{l}{\lambda_g}, \quad W_{qw} \text{ width of quantum well} \tag{13}$$

NUMERICAL ESTIMATE

For an approximate numerical estimate of the above equations, we choose the following values of the parameters involved:

$$E_{BG}(\text{band gap energy}) = 0.625 \, eV; \quad \omega_0 = 10^{15} \, \text{sec}^{-1}, \text{ and } n_1 = 3.5$$

$$F(\omega_o) = \left(\frac{\omega_o^4 \, n_i^4}{\pi \, c^4}\right) = \sim 6.4 \times 10^{19}/\text{cm}^4$$

$$h = 10^{-27} \, ergs.\sec, \quad \lambda_g = \lambda_o = 2 \times 10^{-4} \, cm = 2 \times 10^4 \, \text{Å}$$

$$|\mu_{mn}(1)|^2 = |\mu_{mn}(2)|^2 = \sim 2 \times 10^{-33} \, ergs \, cm^3$$

$$N_1^a = N_2^a = 2 \times 10^{16}/cm^2; \quad T_1 = 1000 \, K, \, T_2 = 300 \, K,$$

$$W_{qw} = 100 \, \text{Å}, \quad l = 1000 \, \text{Å}, \text{ and}$$

$$\bar{l} = (l/\lambda_g) = (1/20).$$

We have $\quad G(\bar{l}) = 24, \quad T_{12}(\bar{l}) \approx 1, \quad F(D) = 4.5 \times 10^{-33},$

and for $T_1 \gg T_2$, $\quad F(B) = e^{-h\omega_0/K_B T_l} = 7 \times 10^{-4}$

as the factor $n(\omega_0, T_2)$ is too small to contribute.

Choosing $\varepsilon_0 = 1, \, \varepsilon_1' = \varepsilon_2' = 12, \quad F(M) = 3.5 \times 10^{-5}.$

The net power transfer between two adjacent 2-dimensional structures (from side 1 to 2) is

$$P_{NF}(aa) = \sim 0.1 \, kW/cm^2$$

We have used plausible (and conservative) values for the various parameters involved. However, because of the high order in many cases (e.g., the fourth power), a change of some values by a factor of 2 can result in order-of-magnitude changes in the calculated power transfer.

Useful energy transfer has been shown to increase dramatically with decreasing gapwidth. As the absorber is removed to a distance from the source, the transverse waves can become the propagating photon mode. Figure 3 indicates the modeled gap dependence ($1/l^4$) of the transverse mode. These effects have been confirmed by computer modeling as well.[21] The curvature at the bottom of the curve indicates the transition into the propagating mode. (The $n^2$ effect, with its $1/l^2$ dependence, may dominate in this transition region; or, it may be small compared to a greater enhancement resulting from the calculated proximity-resonance energy transfer.) Below the $10^{-7}$m gap region, the gap approaches the quantum well width selected and the $1/l^4$ dependence (from eq. 8) begins to roll over (not shown) and approaches the $1/l^2$ dependence of a thick well or a bulk semiconductor. Since the quantum-well result at high gap width is less than that of bulk materials, it will never cross over the bulk semiconductor curve.

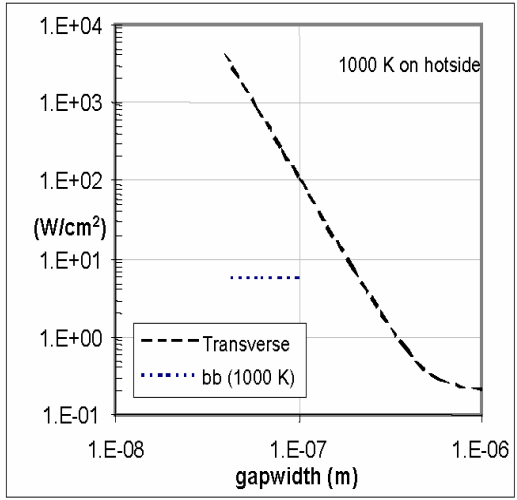

*Figure 3. The gap dependence of the power transferred. A) $1/l^4$ dependence; B)*

Included in the figure is the total radiating power from a 1000 K blackbody. At 0.1 μm, the sum of the proximity resonance energy transfer is >10 times that of the total blackbody radiation. Furthermore, this radiation may be nearly monochromatic rather than the broadband spectrum of the blackbody spectrum (most of which is not useful for photoconversion into electricity). Therefore, all the resonance-mode energy can be useful and significantly enhanced by the use of microgap structures. However, additional energy contributions from proximity coupling of free-carriers and absorption of propagating modes from outside of the quantum wells must be considered in the final accounting of the energy-transfer balance and utility.

CONCLUDING REMARKS

A quantum-mechanical description of a problem that has been explored previously in a more classical setting has indicated that resonance effects can dominate energy transfer between a closely-spaced emitter and detector. Recently, Mulet et al.[22] have indicated highly-enhanced radiative heat transfer between a small particle and a plane surface, which can support resonant-surface waves. Experimental confirmation of the resonant-proximity effects is the current goal.

The coupling between the atoms of the two slabs in this paper arises from the multipolar expansion of the Coulomb interaction between the charges (dipoles) of both sides mediated by the quanta of the electromagnetic field. In the near-field regime, dipole-dipole interaction is realized by the transfer of energy via transverse and longitudinal modes of the electromagnetic field.

The contribution of the longitudinal mode[12] turns out to be of the same order as that of the near-field transverse mode, which involves emission and absorption processes of real photons at resonance in the two-level situation envisioned. For the current case, the speed of light in the appropriate media is modified by the effective refractive index (c --> $c/n_e$). Thus, a factor $n_e^4$ occurs, which also enhances the power transfer. Further enhancement is possible if, instead of a vacuum gap, a dielectric gap is considered with $n_g > 1$. (Enhancement occurs for $n_g$ values up to that of the emitter and collector.) Such effects are particularly important in the near-field region.

The current quantum mechanical resonant-coupling model shows that the non-propagating energy-transfer levels are higher than those obtained by classical models, thus removing a major limitation in thermophotovoltaics. Not only can the efficiency of TPV converters be increased, the radiative efficiency can also be improved because energy that would otherwise be lost (in the form of long-wavelength light or heat) is selectively coupled into a resonant TPV converter. This selectivity further allows a reduction in the temperature of the emitter, while maintaining useful overall system efficiencies. A key feature to remember is that the new energy-transfer mechanism does not only depend on release of the blackbody radiation trapped within the emitter (as does the classical $n^2$ effect). The additional energy source is the non-propagating photon (evanescent wave) modes that are normally dissipated in self-excitation of the emitter atoms. This means that the blackbody law of power emission (which pertains only to the propagating modes) is not violated. We cannot get more power out than we put in. However, we <u>can</u> extract energy more rapidly and more selectively at any emitter temperature. Therefore, for microgap coupling and a given thermal-energy input, the emitter can be kept at a lower temperature and still operate at a higher efficiency than previously possible.

ACKNOWLEDGEMENTS

This paper is based on work supported by The Charles Stark Draper Laboratory, Inc., Cambridge, MA 02139 and is funded in part by the Science for Humanity Trust, Bangalore, 560012, India and the Science for Humanity Trust, Inc, Tucker, GA, USA.